\documentclass{aastex}          %% The manuscript based on AASTeX v5.x
\usepackage{spr-astr-addons}    %% mimicing ASTR journal style
\usepackage{url}
\begin{document}

\title{Distance scale for high-luminosity stars in OB~associations and in  field with {\it Gaia} DR2. Spurious systematic motions }

\shorttitle{Distance scale for high-luminosity stars with {\it
Gaia} DR2}

\shortauthors{Melnik, Dambis}

\author{A.~M. Melnik\altaffilmark{*}} \and
\author{A.~K. Dambis\altaffilmark{}}

\altaffiltext{}{Sternberg Astronomical Institute, Lomonosov Moscow
State University, Universitetskij pr. 13, Moscow 119991, Russia}

\altaffiltext{*}{e-mail: anna@sai.msu.ru}

\begin{abstract}

We calculated the median parallaxes for 47 OB~associations
including at least 10 stars with known {\it Gaia} DR2 parallaxes.
A comparison between trigonometric and photometric parallaxes of
OB~associations reveals a zero-point offset of $\Delta
\varpi=-0.11\pm0.04$ mas indicating that {\it Gaia} DR2 parallaxes
are, on average, underestimated and the distances derived from
them are overestimated. The correction of $\Delta \varpi=-0.11$
mas is consistent with the estimate that Arenou et al. (2018)
obtained for bright stars. An analysis of parallaxes of
OB~associations and high-luminosity field stars confirms our
previous conclusion (Dambis et al. 2001) that the distance scale
for OB~stars established by Blaha and Humphreys (1989) must be
reduced by 10--20\%. Spurious systematic motions of 10--20 km
s$^{-1}$  at the distances of 2--3 kpc from the Sun are found to
arise from the use of the uncorrected {\it Gaia} DR2 parallaxes.
\end{abstract}


\section{Introduction}

The second intermediate {\it Gaia} data release ({\it Gaia} DR2)
includes  high-precision proper motions and parallaxes for 1.3
billion stars \citep{brown2018, lindegren2018a}, which open up new
possibilities for the study of the Galactic structure and kinematics
\citep[][and other papers]{katz2018,fragkoudi2019,carrillo2019,
hunt2019,pettitt2020}.

The  Hipparcos \citep{esa1997} and {\it Gaia} \citep{gaia2016}
satellites have the possibility to measure the absolute parallaxes,
but this capability is susceptible to various instrumental effects,
especially to the basic-angle variations. The basic angle monitor
(BAM) effectively corrects the changes of the basic angle but the
remaining small variations cannot be removed \citep{lindegren2018a}.

Many researchers investigated the zero-point bias, $\Delta \varpi$,
of {\it Gaia} DR2 parallaxes but different studies give different
zero-point corrections. \citet{lindegren2018a} derived $\Delta
\varpi=-0.029\pm0.002$ mas, which means that {\it Gaia} parallaxes
are systematically underestimated and must be increased by  0.029
mas, i.e. distances for all {\it Gaia} DR2 stars must be decreased.
\citet{stassun2018} compared the parallaxes of eclipsing binaries
with {\it Gaia} DR2 parallaxes and found a systematic difference of
$\Delta \varpi = -0.082\pm0.33$ mas. \citet{zinn2019} obtained the
zero-point offset equal to $\Delta \varpi= -0.053 \pm 0.003$ mas
using  stars from the red giant branch. \citet{riess2018} found the
offset to be  $\Delta \varpi = -0.046\pm 0.013$ mas from an analysis
of bright Cepheids. \citet{leung2019} compared spectro-photometric
parallaxes of APOGEE stars to {\it Gaia} DR2 parallaxes and obtained
a zero point bias of $\Delta\varpi = -0.052\pm0.002$ mas. A parallax
correction close to  -0.05 mas is found in many other studies
\citep{yalyalieva2018,schonrich2019}.

Moreover, there is evidence that the zero-point offset depends on the
stellar color and  magnitude \citep{zinn2019, arenou2018, leung2019}.
\citet[][Table 1]{arenou2018} compared {\it Gaia} DR2 parallaxes with
other catalogs and determined the zero-point difference in parallaxes
for different samples of stars. Their analysis reveals a dependence
between the zero-point offset $\Delta \varpi$ and the average G-band
magnitude of stars in the catalog: the brighter the stars the larger
the absolute value of the zero-point offset, $|\Delta \varpi|$.

OB~associations are sparse groups of O- and B-type stars \citep[for
example,][]{ambartsumian1949,blaauw1964}. In this paper we study the
zero-point bias in parallaxes for OB~associations and high-luminosity
field stars with photometric distance scale established by
\citet{blahahumphreys1989}, derive the rotation curve from {\it Gaia}
DR2 data, and study the systematic non-circular motions. Section 2
describes the kinematical data for stars  of OB~associations and
high-luminosity field stars. In Section 3 we compare the photometric
and trigonometric parallaxes, study the distance scale and  the
Galactic rotation curve, presents the systematic motions calculated
for different distance scales. Section 4 discusses the results and
formulates main conclusions.

\section{Data}

The catalog  of  Galactic high-luminosity stars by
\citet{blahahumphreys1989} includes two parts: stars in
OB~associations  and stars scattered  in the field. Both catalogs
present photometric data for  main-sequence O--B2-type stars,
bright giants of spectral types O--B3, and supergiants of all
spectral types. Note that the fraction of red supergiants of
spectral types K and M is only 5\% in both catalogs. The catalog
of stars in OB~associations contains 2209 stars of 91
OB~associations located within $\sim3$ kpc from the Sun. The
catalog of  high-luminosity field stars includes  2492 objects
which do not show the concentration to any groups.
\citet{blahahumphreys1989} derive the distances to OB~associations
and to  field stars, $r_{bh}$, on the basis of their spectral
types and luminosity classes. Stars of both catalogs are massive
young stars and their ages do not exceed 40 Myr
\citep{bressan2012}.

We supplemented the catalogs by \citet{blahahumphreys1989} with
kinematical data for high-luminosity stars. We cross-matched both
catalogs  with the {\it Gaia} DR2 data to search for precise
proper motions and parallaxes, which we found for $\sim 90$\% of
stars. Only 7\% of stars from the list by
\citet{blahahumphreys1989} have line-of-sight velocities, $V_r$,
measured by the {\it Gaia} spectrometer, so here we use the
velocities $V_r$ from the catalog by \citet{barbierbrossat2000},
which are available for 52\% of stars of both catalogs.

In this paper we use the refined sample of stars in OB~associations
and high-luminosity  field stars  which includes only stars with the
re-normalised unit weight errors (RUWE) less than RUWE$<1.4$ and with
the number of visibility periods $n_{vis}>8$
\citep{arenou2018,lindegren2018a,lindegren2018c}. The refined sample
of stars in OB~associations includes 1771 stars with $n_{vis}>8$ and
RUWE$<1.4$; of 219 excluded stars, 174 stars have RUWE$\ge1.4$ and 45
objects have $n_{vis}\le 8$.

We described the catalog of stars of OB~associations in our previous
papers \citep{melnik2017,melnik2020}. Here we give the description of
the second part of the catalog of high-luminosity stars compiled by
\citet{blahahumphreys1989}. Of 2492 field stars 2340 (94\%) are
cross-matched   with {\it Gaia} DR2 catalog and 2319 stars have {\it
Gaia} DR2 proper motions and parallaxes.   We excluded from the full
sample  66 stars with $n_{vis}\le 8$ and 164 stars  with
RUWE$\ge1.4$. Thus, the refined sample of field stars contains 2089
objects. Table~1 (available in the online version of the paper) lists
the kinematic and photometric data for high-luminosity stars in the
field. It presents the name of a star, spectral type, luminosity
class,  color indices $B-V$ and $U-B$, apparent and absolute
magnitudes, $m_V$ and $M_V$, and the $V$-band extinction, $A_V$, that
are adopted from the catalog by \citet{blahahumphreys1989}. We
present the heliocentric distance to the star by
\citet{blahahumphreys1989}, $r_{bh}$, reduced to the short distance
scale, $r=0.8\, r_{bh}$, which is consistent with the
\citet{berdnikov} distance scale for classical Cepheids
\citep{sitnik1996,dambis2001, melnikdambis2009}.  The absolute
magnitudes obtained by \citet{blahahumphreys1989}, $M_{V(BH)}$, were
converted to  the short distance scale $M_V$=$M_{V(BH)}$+$\Delta m$,
where $\Delta m=-5\log 0.8=0.485^m$. Table~1 also lists  {\it Gaia}
DR2 data: equatorial coordinates, $\alpha$ and $\delta$, of the star;
its Galactic coordinates, $l$ and $b$; the $G$-band magnitude; the
parallax, $\varpi$; proper-motion components along $l$- and
$b$-directions, $\mu_l$ and $\mu_b$, and their errors,
$\varepsilon_{\varpi}$, $\varepsilon_{\mu_l}$ and
$\varepsilon_{\mu_b}$; the error RUWE and the number of visibility
periods, $n_{vis}$. Table~1 also gives the line-of-sight velocities,
$V_r$, and their errors, $\varepsilon_{vr}$, taken from the catalog
by \citet{barbierbrossat2000}.

%----------------------- Figure 1  --------------------------
\begin{figure}
\resizebox{\hsize}{!}{\includegraphics{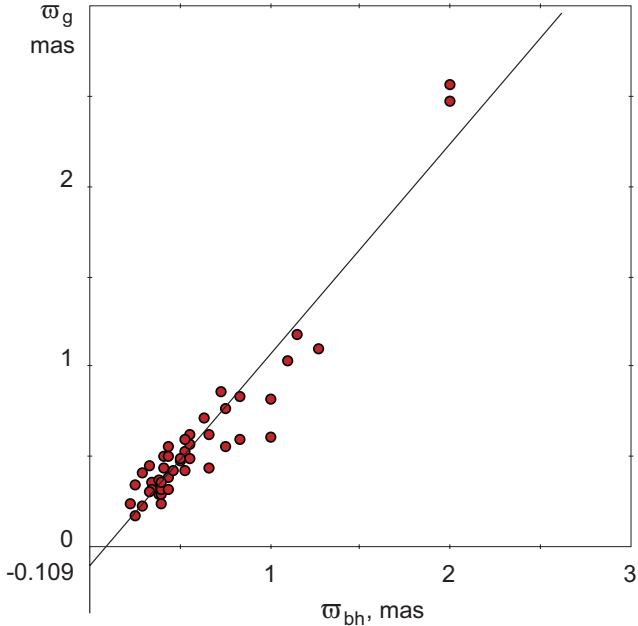}}
\caption{Comparison between trigonometric, $\varpi_g$, and
photometric, $\varpi_{bh}$, parallaxes of 47 OB~associations. The
straight line shows the linear dependence between these
quantities, $\varpi_g=1.16\,\varpi_{bh}-0.11$ mas. We can see that
it does not pass through the origin but crosses the vertical axis
at negative values, which is indicative of a systematic excess of
photometric parallaxes over trigonometric parallaxes.
Consequently, {\it Gaia} DR2 parallaxes must be increased, whereas
the corresponding distances to OB~associations must be reduced.}
\label{piob}
\end{figure}
%-------------------------------------------------------------

%----------------------- Figure 2  --------------------------
\begin{figure*}
\resizebox{\hsize}{!}{\includegraphics{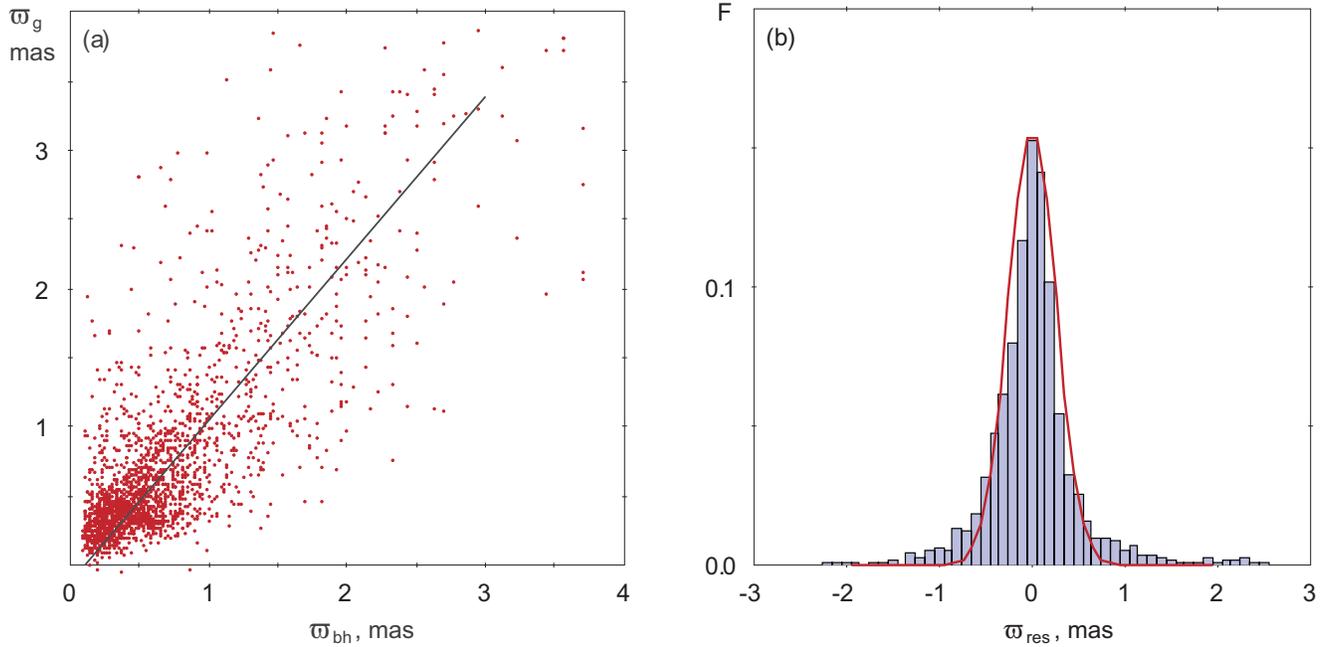}} \caption{(a)
Trigonometric, $\varpi_g$, parallaxes plotted as a function of
photometric, $\varpi_{bh}$, parallaxes for 2089 high-luminosity field
stars. The straight line shows the linear dependence between these
quantities (Eq.~\ref{pig2}). The most probable value of $k_p$ appears
to be $1.16\pm0.01$. (b) Distribution of the residual parallaxes,
$\varpi_{res}$ (Eq.~\ref{pi_res}). The solid curve shows the Gaussian
distribution with the standard deviation of $\sigma=0.26$ mas, which
fits well the observed distribution in the central
$\pm2\sigma$-interval. } \label{pifd}
\end{figure*}
%-------------------------------------------------------------

%----------------------- Figure 3  --------------------------
\begin{figure}
\resizebox{\hsize}{!}{\includegraphics{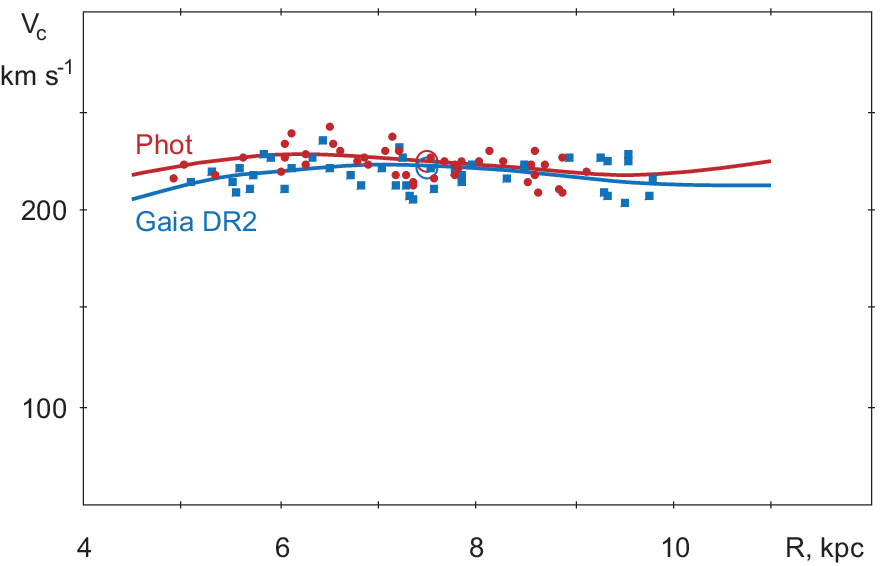}} \caption{Two
rotation curves of the Galactic disk derived with the use of
photometric  and trigonometric  distances to OB~associations. The
large circle indicates the position of the Sun. The azimuthal
velocities of OB~associations calculated with the use of photometric
and trigonometric  distances are indicated by circles and squares,
respectively. } \label{rot}\end{figure}
%-------------------------------------------------------------

\section{Results}

\subsection{Zero-point bias of the distance scale to OB~associations}

The trigonometric parallaxes to OB~associations, $\varpi_g$, are
determined as the median values of {\it Gaia} DR2 parallaxes of
their member stars. We selected 47 OB~associations including more
than 9 stars with known {\it Gaia} DR2 parallaxes. The photometric
parallaxes are calculated as inverse values of the photometric
distances, $\varpi_{bh}=1/r_{bh}$, to OB~associations derived by
\citet{blahahumphreys1989}. The least-squares solution of 47
linear equations:

\begin{equation}
\varpi_g=k_p\,\varpi_{bh}+ \Delta \varpi \label{pi}
\end{equation}

\noindent gives the most probable values of the coefficients equal
to $k_p=1.159\pm0.055$ and $\Delta \varpi=-0.109\pm0.039$ mas:

\begin{equation}
\varpi_g=(1.159\pm0.055)\,\varpi_{bh}-(0.109\pm0.039)\; \textrm{mas}.
\label{pi1}
\end{equation}

\noindent The root-mean-square deviation from the linear
dependence appears to be 0.14 mas.

Figure~\ref{piob} shows the distribution of trigonometric,
$\varpi_g$, and photometric, $\varpi_{bh}$, parallaxes of 47
OB~associations. The  linear dependence between them determined by
Eq.~\ref{pi1} is shown by the straight line. We can see that the
straight line does not pass through the origin but crosses the
vertical axis at the negative value of trigonometric parallaxes
indicating a systematic excess of photometric parallaxes over
trigonometric parallaxes. The systematic offset of trigonometric
parallaxes over photometric parallaxes  is determined mainly by
distant objects located beyond  1 kpc from the Sun, $r>1$ kpc. On
the contrary, the coefficient of the distance scale  is determined
mainly by nearby objects, $r<1$ kpc.

A comparison of trigonometric and photometric parallaxes requires
some caution: they have different distributions of errors which can
give rise to systematic errors \citep{luri2018}. Note that the random
error in photometric distances to OB~associations (without the
allowance for the uncertainty of the zero-point of the distance
scale) is, on average, 6\% \citep{melnikdambis2009}. We simulated the
distribution of observational errors in photometric distances and
trigonometric parallaxes and calculated the biases in parameters
$k_p$ and $\Delta \varpi$ caused by such errors. The distance
modulus, $m_V-M_V$, of an object without correction for extinction is
determined by the relation:

\begin{equation}
 DM= 5\lg r +10, \label{dm}
\end{equation}

\noindent where $r$ is in kpc. Here we assume that the distance
moduli of stars of OB association are obtained with a random error of
0.5$^m$, so the uncertainty in the distance modulus of an
OB~association must be equal to:

\begin{equation}
\sigma_m=0.5^m/\sqrt{n_t}, \label{e_r}
\end{equation}

\noindent where $n_t$ is the number of stars of the OB~association
with known photometry (see Table~3). The total errors in {\it Gaia}
DR2 parallaxes are believed to be given  by formula:

\begin{equation}
\sigma_p=\sqrt{k^2\sigma_i^2+\sigma_s^2}, \label{e_par}
\end{equation}

\noindent where $\sigma_i$ is the formal uncertainty in a stellar
parallax (possibly underestimated)  and $\sigma_s$ is the systematic
error in parallaxes. The factor $k$ and systematic error $\sigma_s$
take different values for bright ($G<13^m$) and faint ($G>13^m$)
stars. Note that the average $G$ magnitude of stars of OB~association
from the catalog by \citet{blahahumphreys1989} is
$\overline{G}=8.5^m$. We therefore adopted the $k$ and $\sigma_s$
values  equal to $k=1.08$ and $\sigma_s=0.021$ mas, respectively
\citep{arenou2018,lindegren2018a,lindegren2018b}.

We suppose that  true  values of  $\varpi_g$ and
$\varpi_{bh}=1/r_{bh}$ are connected though Eq.~\ref{pi1}. As we do
not know true values we use observational values instead of them.
Here we implicitly assume that  the distributions of true and wrong
values  are nearly the same. We simulated the observational errors in
distance moduli $DM$($r_{bh}$) (Eq~\ref{dm}) of OB associations and
in parallaxes $\varpi_g$  by adding normally distributed values with
the standard deviations determined by Eqs~\ref{e_r} and \ref{e_par},
respectively. Then we derived the parameters $k_p$ and $\Delta
\varpi$ from  'wrong' data. The calculated values, $k_{p}'$ and
$\Delta \varpi'$, of the parameters appear to be slightly different
from the true ones, $k_{p}^0$ and $\Delta \varpi^0$.  We modelled
10$^3$ observational samples and estimated the average values of
systematic corrections:

\begin{equation}
k_p'-k_p^0=-0.008, \label{sys_kp}
\end{equation}

\begin{equation}
\Delta \varpi'-\Delta \varpi^0=0.004, \;\textrm{mas}
\label{sys_kp}
\end{equation}

\noindent which must be subtracted from  the calculated earlier
values $k_{p}=1.159$ and $\Delta \varpi=-0.109$ mas (Eq.~\ref{pi1}).
So the unbiased, 'true', values of the parameters relating
trigonometric and photometric parallaxes are:

\begin{equation}
\varpi_g=(1.167\pm0.055)\,\varpi_{bh}-(0.113\pm0.039)\;
\textrm{mas}. \label{picor}
\end{equation}

The fact that $k_p>1$ indicates that the distance scale
established by \citet{blahahumphreys1989} must be reduced,
$r=k_d\, r_{bh}$,  by the factor $k_d=1/k_p$ equal to
$k_d=0.86^\pm0.04$.

Thus, a comparison of trigonometric and photometric parallaxes of
OB~associations suggests that {\it Gaia} DR2 parallaxes have a
systematic bias of $\Delta \varpi=-0.11\pm0.04$ mas and the
distance scale established by \citet{blahahumphreys1989} requires
a reduction by 10--18\%.

\subsection{Distance scale to  high-luminosity field stars}

The catalog of high-luminosity field stars by
\citet{blahahumphreys1989} includes  2089 objects with reliable {\it
Gaia} DR2  parallaxes, $\varpi_g$. Within 2 kpc from the Sun, {\it
Gaia} DR2 trigonometric distances have formally higher precision than
the photometric distances of individual stars determined from the
color-magnitude calibrations which are accurate to  $\sim0.5^m$ in
terms of distance modulus, $m_V-M_V$. The median heliocentric
distance of field stars  is 1.8 kpc, so we consider them all without
selecting objects with the most precise {\it Gaia} DR2 parallaxes.
Figure~\ref{pifd}(a) shows the distribution of trigonometric,
$\varpi_g$,  and photometric, $\varpi_{bh}$, parallaxes of young
field stars. We can see that distant objects with small parallaxes
form a wide cloud of points near the origin. We cannot determine the
systematic bias in parallaxes in this case so we just adopt a fixed
parallax correction of $\Delta\varpi=-0.113$ mas and introduce it
into the equations:

\begin{equation}
\varpi_g-\Delta\varpi=k_p\,\varpi_{bh} \label{pieq}.
\end{equation}

\noindent The parallax correction of $\Delta\varpi=-0.113$ mas was
derived for the other part of the catalog of high-luminosity
stars, namely for stars located in OB~associations. As the
distribution of deviations is quite sampled in this case, we use a
$4\sigma$ criteria instead of $3\sigma$ one to exclude the
outliers. The least-squares solution of 2038 linear equations with
respect to the distance scale coefficient, $k_p$, gives us its
most probable value:

\begin{equation}
\varpi_g-\Delta\varpi=(1.164\pm0.010)\,\varpi_{bh}, \label{pig2}
\end{equation}

\noindent which defines the slope of the straight line in
Figure~\ref{pifd}(a).

The photometric parallaxes of field stars are determined with large
and asymmetrically distributed errors, so the calculated value of
$k_p$ can be shifted with respect to the true one. We assumed that
the distance moduli of field stars are determined with the random
error of 0.5$^m$ and estimated the bias in distance scale coefficient
$k_p$ using  the method described in section 2.1. The systematic
shift appears to be:

\begin{equation}
k_p'-k_p^0=-0.072, \label{sys_kpf}
\end{equation}

\noindent so the unbiased  value of $k_p$ is $k_p=1.236$.

The value of $k_p=1.24.01$ indicates that the distances to young
field stars derived by \citet{blahahumphreys1989} must be corrected
by a factor $1/k_p=0.81\pm0.01$, i.~e. must be shrunk by $19\pm1\%$.
The root-mean-square deviation of stellar parallaxes from the linear
dependence defined by Eq.~\ref{pig2}   amounts to 0.5 mas.

Figure~\ref{pifd}(b) shows the distribution of the residual values
of parallaxes:

\begin{equation}
\varpi_{res}=\varpi_g-k_p\,\varpi_{bh}-\Delta\varpi
\label{pi_res}\end{equation}

\noindent   We approximated the distribution of residual
parallaxes, $\varpi_{res}$, by a Gauss distribution with  the
standard deviation of $\sigma=0.26$ mas, which fits well the
observed distribution in the central $\pm2\sigma$-interval. It
means that the standard deviation of distance moduli of field
stars is $\sim0.5^m$, which is to be expected  for distances
derived from spectral-luminosity class calibrations in the V-band.
However, there are a lot of stars with large residual parallaxes,
$\varpi_{res}$, which show up as broad wings in the distribution
and are  due to gross errors in both photometric and trigonometric
parallaxes.

\subsection{Galactic rotation curve}

We compared the parameters of the rotation curve derived with the use
of trigonometric and  photometric distances to OB~associations.
Table~2 lists the Galactic coordinates, $l$ and $b$, as well as
trigonometric and  photometric distances, $r_{ph}$ and $r_g$, to 66
OB~associations from the catalog by \citet{blahahumphreys1989}
including at least  5 stars with known {\it Gaia} DR2 parallaxes or
at least 5 stars with known line-of-sight velocities from the catalog
by \citet{barbierbrossat2000}. These photometric distances, $r_{ph}$,
are calculated for the short distance scale: $r_{ph}=0.8\,r_{bh}$
\citep{sitnik1996,dambis2001, melnikdambis2009}. Trigonometric
distances are determined from {\it Gaia} DR2 parallaxes:
$r_g=1/\varpi$. Table~2 also presents the median {\it Gaia} DR2
proper motions of stars of OB~associations, $\mu_l$ and $\mu_b$, and
their uncertainties, $\varepsilon_{\mu l}$ and $\varepsilon_{\mu b}$;
median line-of-sight velocities, $V_r$, and their uncertainties,
$\varepsilon_{vr}$; the number of stars of OB~associations with known
{\it Gaia} DR2 parallaxes (and consequently proper motions), $n_\mu$;
the number of stars with known line-of-sight velocity, $n_{vr}$, from
the catalog by \citet{barbierbrossat2000};  as well as   the total
number of stars of the association with known photometry, $n_t$. The
uncertainties $\varepsilon_{\mu l}$, $\varepsilon_{\mu b}$ and
$\varepsilon_{vr}$ are calculated as half the size of the central
interval containing 67\% of values of $\mu_l$, $\mu_b$ and $V_r$ in
the OB~association, respectively.

%------------------------------------------------------------------------------
\addtocounter{table}{+1}
%------------------------------------------------------------------------------

\begin{table*}
\small \caption{Distances, proper motions and line-of-sight
velocities for OB~associations}
 \begin{tabular}{lcccccccccr}
 \small
 \\[-7pt] \hline\\[-7pt]
 Association & $l$ & $b$ & $r_{ph}$ & $r_g$ & $n_t$ & $\mu_l$ & $\mu_b$ & $n_{\mu}$ & $V_r$  & $n_{vr}$   \\
      &  deg. & deg. & kpc     & kpc &                & mas yr$^{-1}$ & mas yr$^{-1}$ & & km s$^{-1}$&\\
  \\[-7pt] \hline\\[-7pt]
SGR OB5   &   0.04 &  -1.16 &   2.42 &   2.23 &    30 &     -1.809$\pm$0.226 &     -0.810$\pm$0.120 &     27 &   -15.0$\pm$13.4 &    2 \\
SGR OB1   &   7.54 &  -0.77 &   1.26 &   1.40 &    65 &     -1.162$\pm$0.062 &     -1.375$\pm$0.123 &     47 &   -10.0$\pm$2.0 &    37 \\
SGR OB4   &  12.11 &  -0.96 &   1.92 &   1.98 &    15 &     -1.204$\pm$0.103 &     -1.046$\pm$0.139 &     14 &     3.5$\pm$3.6 &     9 \\
SER OB1   &  16.71 &   0.07 &   1.53 &   1.87 &    43 &     -1.325$\pm$0.098 &     -0.839$\pm$0.098 &     33 &    -5.0$\pm$4.9 &    17 \\
SCT OB3   &  17.30 &  -0.73 &   1.33 &   1.91 &    10 &     -2.504$\pm$0.166 &     -0.686$\pm$0.114 &      6 &     3.3$\pm$6.0 &     8 \\
SER OB2   &  18.21 &   1.63 &   1.60 &   2.07 &    18 &     -2.120$\pm$0.156 &     -0.517$\pm$0.082 &     16 &    -4.0$\pm$5.5 &     7 \\
SCT OB2   &  23.17 &  -0.54 &   0.80 &   1.63 &    13 &     -2.194$\pm$0.327 &     -0.949$\pm$0.093 &     10 &   -11.0$\pm$8.2 &     6 \\
VUL OB1   &  60.30 &   0.12 &   1.60 &   2.03 &    27 &     -5.243$\pm$0.264 &     -0.665$\pm$0.305 &     15 &     5.8$\pm$4.6 &     8 \\
VUL OB4   &  60.63 &  -1.22 &   0.80 &   2.12 &     9 &     -4.620$\pm$0.612 &     -1.392$\pm$0.276 &      6 &    -2.9$\pm$4.3 &     3 \\
CYG OB3   &  72.76 &   2.04 &   1.83 &   1.96 &    40 &     -7.029$\pm$0.138 &     -0.737$\pm$0.085 &     32 &   -10.0$\pm$1.8 &    29 \\
CYG OB1   &  75.84 &   1.12 &   1.46 &   1.78 &    71 &     -6.213$\pm$0.115 &     -0.638$\pm$0.066 &     62 &   -13.5$\pm$1.5 &    34 \\
CYG OB9   &  77.81 &   1.80 &   0.96 &   1.68 &    32 &     -5.977$\pm$0.161 &     -0.650$\pm$0.157 &     22 &   -19.5$\pm$2.8 &    10 \\
CYG OB8   &  77.92 &   3.36 &   1.83 &   1.78 &    21 &     -6.034$\pm$0.122 &      0.479$\pm$0.245 &     20 &   -21.0$\pm$3.7 &     9 \\
CYG OB2   &  80.27 &   0.88 &   1.46 &   1.62 &    15 &     -4.845$\pm$0.128 &     -0.359$\pm$0.027 &      9 &                 &     0 \\
CYG OB7   &  88.98 &   0.03 &   0.63 &   0.91 &    29 &     -2.030$\pm$0.780 &     -0.961$\pm$0.159 &     22 &    -9.4$\pm$2.0 &    21 \\
CEP OB2   & 102.02 &   4.69 &   0.73 &   0.97 &    56 &     -3.663$\pm$0.228 &     -0.580$\pm$0.204 &     45 &   -17.0$\pm$1.1 &    36 \\
CEP OB1   & 104.20 &  -0.94 &   2.78 &   4.32 &    58 &     -4.426$\pm$0.114 &     -0.608$\pm$0.058 &     44 &   -58.2$\pm$1.8 &    17 \\
CEP OB5   & 108.50 &  -2.69 &   1.67 &   3.43 &     6 &     -3.392$\pm$0.210 &     -0.969$\pm$0.210 &      6 &   -48.7$\pm$20.6 &     2 \\
CAS OB2   & 111.99 &  -0.00 &   2.10 &   3.42 &    41 &     -3.923$\pm$0.144 &     -0.642$\pm$0.119 &     30 &   -50.1$\pm$4.2 &     7 \\
CEP OB3   & 110.71 &   3.13 &   0.70 &   0.85 &    25 &     -1.891$\pm$0.171 &     -1.115$\pm$0.143 &     18 &   -22.9$\pm$0.9 &    18 \\
CAS OB5   & 116.09 &  -0.50 &   2.01 &   3.46 &    52 &     -3.471$\pm$0.058 &     -0.941$\pm$0.058 &     45 &   -45.8$\pm$1.8 &    16 \\
CEP OB4   & 118.21 &   5.25 &   0.66 &   1.06 &     7 &     -1.957$\pm$0.029 &     -1.144$\pm$0.154 &      7 &   -24.0         &     1 \\
CAS OB4   & 120.05 &  -0.30 &   2.30 &   2.84 &    27 &     -2.923$\pm$0.147 &     -0.535$\pm$0.099 &     24 &   -37.0$\pm$3.3 &     7 \\
CAS OB14  & 120.36 &   0.74 &   0.88 &   1.51 &     8 &     -1.630$\pm$0.384 &     -0.882$\pm$0.266 &      6 &   -15.0$\pm$3.5 &     4 \\
CAS OB7   & 122.98 &   1.22 &   2.01 &   3.10 &    39 &     -2.328$\pm$0.067 &     -0.387$\pm$0.044 &     35 &   -50.0$\pm$0.5 &     4 \\
CAS OB1   & 124.73 &  -1.73 &   2.01 &   2.31 &    11 &     -1.345$\pm$0.257 &     -1.076$\pm$0.137 &      7 &   -42.0$\pm$1.1 &     5 \\
CAS OB8   & 129.16 &  -1.06 &   2.30 &   3.15 &    43 &     -0.996$\pm$0.025 &     -0.498$\pm$0.031 &     41 &   -34.6$\pm$2.6 &    14 \\
PER OB1   & 134.70 &  -3.14 &   1.83 &   2.59 &   163 &     -0.120$\pm$0.035 &     -1.188$\pm$0.029 &    150 &   -43.2$\pm$0.8 &    80 \\
CAS OB6   & 134.95 &   0.72 &   1.75 &   2.36 &    45 &     -0.241$\pm$0.104 &     -0.723$\pm$0.132 &     29 &   -42.6$\pm$2.3 &    12 \\
CAM OB1   & 141.08 &   0.89 &   0.80 &   1.22 &    50 &      0.232$\pm$0.135 &     -1.118$\pm$0.122 &     41 &   -11.0$\pm$1.7 &    30 \\
CAM OB3   & 146.97 &   2.85 &   2.65 &   5.48 &     8 &     -0.032$\pm$0.029 &      0.127$\pm$0.104 &      6 &   -27.6$\pm$11.1 &     3 \\
PER OB2   & 160.22 & -16.55 &   0.32 &   0.41 &     7 &      4.905$\pm$0.448 &     -0.633$\pm$0.684 &      5 &    21.2$\pm$1.7 &     7 \\
AUR OB1   & 173.83 &   0.14 &   1.06 &   1.80 &    36 &      2.621$\pm$0.144 &     -1.806$\pm$0.109 &     31 &    -1.9$\pm$2.7 &    26 \\
ORI OB1   & 206.90 & -17.71 &   0.40 &   0.39 &    68 &      0.928$\pm$0.212 &      0.633$\pm$0.126 &     54 &    25.4$\pm$1.0 &    62 \\
 \\[-7pt] \hline\\[-7pt]
\end{tabular}
\end{table*}
%------------------------------------------------------------------------------
\addtocounter{table}{-1}
%------------------------------------------------------------------------------
\begin{table*}
\small \caption{continued}
 \begin{tabular}{lcccccccccr}
\footnotesize
 \\[-7pt] \hline\\[-7pt]
 Association & $l$ & $b$ & $r_{ph}$ & $r_g$ & $n_t$ & $\mu_l$ & $\mu_b$ & $n_{\mu}$ & $V_r$  & $n_{vr}$   \\
      &  deg. & deg. & kpc     & kpc &                & mas yr$^{-1}$ & mas yr$^{-1}$ & & km s$^{-1}$&\\
  \\[-7pt] \hline\\[-7pt]
AUR OB2   & 173.33 &  -0.16 &   2.42 &   3.24 &    20 &      1.928$\pm$0.173 &     -1.188$\pm$0.072 &     16 &    -2.6$\pm$2.5 &     4 \\
NGC 1893  & 173.60 &  -1.70 &   2.90 &   3.36 &    10 &      1.035$\pm$0.176 &     -1.284$\pm$0.062 &      6 &                 &     0 \\
GEM OB1   & 188.96 &   2.22 &   1.21 &   2.26 &    40 &      1.900$\pm$0.113 &     -0.713$\pm$0.076 &     35 &    16.0$\pm$1.2 &    18 \\
MON OB1   & 202.08 &   1.08 &   0.58 &   0.87 &     7 &      1.348$\pm$0.347 &     -2.315$\pm$0.419 &      6 &    23.4$\pm$4.9 &     7 \\
MON OB2   & 207.35 &  -1.60 &   1.21 &   1.60 &    31 &     -0.782$\pm$0.119 &     -1.663$\pm$0.146 &     23 &    23.0$\pm$2.3 &    25 \\
CMA OB1   & 224.58 &  -1.56 &   1.06 &   1.30 &    17 &     -2.781$\pm$0.224 &     -2.861$\pm$0.280 &     16 &    34.3$\pm$5.7 &     8 \\
COLL 121  & 238.42 &  -8.41 &   0.55 &   0.65 &    13 &     -5.626$\pm$0.281 &     -1.192$\pm$0.227 &      8 &    29.6$\pm$2.2 &    10 \\
NGC 2362  & 237.82 &  -5.96 &   1.21 &   1.46 &     8 &     -4.024$\pm$0.540 &     -0.964$\pm$0.153 &      3 &    18.0$\pm$6.3 &     5 \\
NGC 2439  & 245.27 &  -4.08 &   3.50 &   4.12 &    23 &     -3.910$\pm$0.077 &     -0.579$\pm$0.041 &     22 &    62.7         &     1 \\
PUP OB1   & 243.53 &   0.16 &   2.01 &   4.22 &    22 &     -3.791$\pm$0.065 &     -0.857$\pm$0.061 &     16 &    77.0         &     1 \\
PUP OB2   & 244.61 &   0.58 &   3.18 &   5.74 &    13 &     -3.440$\pm$0.345 &     -0.652$\pm$0.228 &      9 &                 &     0 \\
COLL 140  & 244.42 &  -7.33 &   0.29 &   0.42 &     6 &     -6.974$\pm$0.287 &     -4.512$\pm$0.598 &      6 &    10.3$\pm$3.0 &     5 \\
VELA OB2  & 262.05 &  -8.52 &   0.40 &   0.40 &    13 &     -9.620$\pm$0.605 &     -0.127$\pm$0.380 &      9 &    24.0$\pm$2.7 &    13 \\
VELA OB1  & 264.83 &  -1.41 &   1.46 &   2.06 &    46 &     -6.978$\pm$0.092 &     -1.461$\pm$0.078 &     43 &    23.0$\pm$1.0 &    18 \\
CAR OB1   & 286.45 &  -0.46 &   2.01 &   2.99 &   126 &     -7.581$\pm$0.079 &     -0.731$\pm$0.033 &    101 &    -5.0$\pm$1.3 &    39 \\
TR 16     & 287.25 &  -0.25 &   2.10 &   2.72 &    18 &     -7.300$\pm$0.099 &     -1.067$\pm$0.077 &     14 &    -1.0$\pm$1.5 &     5 \\
COLL 228  & 287.57 &  -0.98 &   2.01 &   3.10 &    15 &     -6.556$\pm$0.105 &     -1.538$\pm$0.065 &     13 &   -13.0$\pm$3.0 &     9 \\
CAR OB2   & 290.39 &   0.12 &   1.83 &   3.10 &    59 &     -6.467$\pm$0.064 &     -1.152$\pm$0.044 &     48 &    -8.2$\pm$1.8 &    22 \\
CRU OB1   & 294.87 &  -1.06 &   2.01 &   2.82 &    75 &     -6.161$\pm$0.054 &     -1.034$\pm$0.033 &     65 &    -5.3$\pm$1.5 &    33 \\
NGC 3766  & 294.12 &  -0.02 &   1.53 &   2.40 &    11 &     -6.649$\pm$0.028 &     -1.038$\pm$0.031 &     10 &   -15.6$\pm$0.5 &     2 \\
CEN OB1   & 304.14 &   1.44 &   1.92 &   2.27 &   103 &     -4.712$\pm$0.061 &     -1.090$\pm$0.034 &     85 &   -19.0$\pm$2.6 &    32 \\
NGC 5606  & 314.87 &   0.99 &   1.53 &   2.50 &     5 &     -5.622$\pm$0.040 &     -0.862$\pm$0.059 &      5 &   -37.8$\pm$1.0 &     3 \\
PIS 20    & 320.39 &  -1.48 &   3.18 &   3.38 &     6 &     -4.845$\pm$0.095 &     -0.694$\pm$0.097 &      5 &   -49.0         &     1 \\
NOR OB1   & 328.05 &  -0.92 &   2.78 &   2.52 &     8 &     -3.946$\pm$0.094 &     -0.699$\pm$0.110 &      7 &   -35.6$\pm$2.7 &     6 \\
NGC 6067  & 329.71 &  -2.18 &   1.67 &   2.03 &     9 &     -3.250$\pm$0.071 &     -0.454$\pm$0.075 &      8 &   -40.0$\pm$0.9 &     8 \\
R 103     & 332.36 &  -0.74 &   3.18 &   2.93 &    33 &     -3.843$\pm$0.273 &     -0.827$\pm$0.086 &     26 &   -47.5$\pm$8.2 &    10 \\
ARA OB1B  & 337.95 &  -0.85 &   2.78 &   2.47 &    21 &     -2.511$\pm$0.092 &     -0.936$\pm$0.081 &     19 &   -34.7$\pm$3.4 &     9 \\
ARA OB1A  & 337.68 &  -0.92 &   1.10 &   1.16 &    53 &     -2.064$\pm$0.137 &     -2.629$\pm$0.231 &     42 &   -36.3$\pm$7.3 &     8 \\
NGC 6204  & 338.34 &  -1.16 &   2.20 &   2.83 &    13 &     -2.079$\pm$0.038 &     -0.598$\pm$0.062 &      5 &   -51.0$\pm$2.6 &     5 \\
SCO OB1   & 343.72 &   1.37 &   1.53 &   1.67 &    73 &     -1.823$\pm$0.038 &     -0.806$\pm$0.041 &     66 &   -28.8$\pm$2.9 &    28 \\
SCO OB2   & 351.29 &  19.02 &   0.13 &   0.15 &    10 &    -23.339$\pm$0.791 &     -8.307$\pm$0.132 &      4 &    -4.1$\pm$0.7 &    10 \\
SCO OB4   & 352.64 &   3.23 &   0.96 &   1.20 &    11 &     -0.633$\pm$0.189 &     -2.672$\pm$0.083 &     10 &     3.0$\pm$2.4 &     7 \\
 \\[-7pt] \hline\\[-7pt]
\end{tabular}
\end{table*}

We determined the parameters of the rotation curve and the motion of
the Sun towards the apex by solving the set of Bottlinger equations
for  line-of-light velocities and proper motions of associations:

\begin{equation}
\begin{array}{c}
4.74 r \mu_l (\cos b)^{-1} =R_0(\Omega-\Omega_0)\cos l - \Omega \,r \cos b \\
\\
+u_0\sin l -v_0\cos l. \label{mu}
\end{array}
\end{equation}

\begin{equation}
\begin{array}{c}
V_r=R_0(\Omega-\Omega_0)\sin l \cos b  \\
\\
-u_0 \cos l \cos b-v_0\sin l \cos b-w_0 \sin b,
\label{vr}\end{array}
\end{equation}

\noindent where the coefficient  $4.74\times r$ transforms proper
motions in units mas yr$^{-1}$ into  tangential velocities in  km
s$^{-1}$;  the factor $(\cos b)^{-1}$ in the left part of
Eq.~\ref{mu} converts local proper motions $\mu_l$ measured in the
direction parallel to the Galactic plane into the motions in the
Galactic plane; $\Omega$ and $\Omega_0$ are the angular velocities of
the differential circular rotation of the Galactic disk determined at
the Galactocentric distances, $R$, of the center of the association
and at the solar distance, $R_0$; the components of the solar motion
towards the apex, $u_0$, $v_0$ and $w_0$, are directed toward the
Galactic center, in the sense of Galactic rotation and toward the
Galactic North Pole, respectively.

We expand  the difference  $\Omega - \Omega_0$   into a power series
in $(R-R_0)$:

\begin{equation}
\Omega- \Omega_0=\Omega'_0 (R-R_0)+0.5\Omega''_0 (R-R_0)^2,
\end{equation}

\noindent where $\Omega'_0$ and $\Omega''_0$ are the first and second
derivatives taken at the solar distance,  $R_0$. So the Eqs \ref{mu}
and \ref{vr} can be rewriten in the following way:

\begin{equation}
\begin{array}{c}
4.74 r \mu_l (\cos b)^{-1}=- \Omega_0 r \cos b +u_0\sin l -v_0\cos l\\
\\
+\Omega'_0 (R-R_0)(R_0 \cos l-r \cos b)   \\
\\
+0.5\Omega''_0 (R-R_0)^2 \,(R_0 \cos l -r \cos b),\\
\label{mu1}
\end{array}
\end{equation}

\begin{equation}
\begin{array}{c}
V_r=-u_0 \cos l \cos b-v_0\sin l \cos b-w_0 \sin b\\
\\
+\Omega'_0 R_0(R-R_0)\sin l \cos b \\
\\
+0.5\Omega''_0 R_0 (R-R_0)^2\sin l \cos b,  \\
\label{vr1}
\end{array}
\end{equation}

We solve the sets of equations for proper motions (\ref{mu1}) and
line-of-sight velocities (\ref{vr1}) jointly applying the weight
factors, $p_{vl}$ and $p_{vr}$, which take into account the
observational errors,  systematic error in proper motions,
$\sigma_{\mu s}$, and "cosmic" velocity dispersion:

\begin{equation}
p_{vl} =(\sigma_0^2+(4.74\,r \varepsilon_{\mu t})^2)^{-1/2},
\label{pvl}\end{equation}

\begin{equation}
p_{vr} =(\sigma_0^2+\varepsilon^2_{vr})^{-1/2}, \label{pvr}
\end{equation}

\noindent where the total error in proper motion $\mu_l$ is

\begin{equation}
\varepsilon_{\mu  t} =\sqrt{\varepsilon_{\mu l}^2+\sigma_{\mu s}},
\label{et}\end{equation}

\noindent We assume the "cosmic" dispersion  to be $\sigma_0=7.0$ km
s$^{-1}$, which nearly coincide with the root-mean-squared deviation
of  velocities of OB~associations from the Galactic rotation curve
\citep[for more details,][]{dambis1995, dambis2001,melnikdambis2009}.
The  systematic error in {\it Gaia} DR2 proper motions is supposed to
be $\sigma_{\mu s}=0.055$ mas yr$^{-1}$
\citep{arenou2018,lindegren2018a,lindegren2018b}.

We adopted a solar Galactocentric distance to be of $R_0=7.5$ kpc
\citep{glushkova1998, nikiforov2004, feast2008, groenewegen2008,
reid2009b, dambis2013, francis2014, boehle2016, branham2017}.

Table~3 lists the parameters of the Galactic rotation curve,
$\Omega_0$, $\Omega'_0$ and $\Omega''_0$, and the solar motion
towards the apex, $u_0$ and $v_0$, calculated for two sets of
distances to OB~associations:  photometric  and trigonometric ones.
We  use the median proper motions and line-of-sight velocities
derived from kinematical data of at least 5 member stars, $n_\mu\ge
5$ or $n_{vr}\ge 5$, respectively. Given that Gaia DR2 proper motions
and line-of-sight velocities from the catalog by
\citet{barbierbrossat2000} are available  for 90\% and 52\% of
OB~association stars, respectively, adopting the minimal number of
stars with the corresponding data equal to 5, $n_\mu\ge 5$ or
$n_{vr}\ge 5$, gives us noticeably different numbers of conditional
equations for proper motions and line-of-sight velocities: 64 and 50,
respectively. Table~3 also lists the value of the Oort constant,
$A=-0.5\Omega'_0 R_0$, and the standard deviation of the velocities
from the rotation curve, $\sigma_0$. It also gives the number of
conditional equations  for proper motions (Eq.~\ref{mu1}) and
line-of-sight velocities (Eq.~\ref{vr1}) in the form: $N_\mu+N_{vr}$.
OB~associations and young field stars are located close to the
Galactic plane and  the component of the solar velocity, $w_0$, is
poorly determined from the solution of equations for line-of-sight
velocities,    and we therefore adopted the value of $w_0=7.0$ km
s$^{-1}$.

%-------------------  Table 3  ---------------------------------
\begin{table*}
\centering \caption{Parameters of the Galactic rotation curve and
the solar motion towards the apex}
 \begin{tabular}{lcccccccc}
 \\[-7pt] \hline\\[-7pt]
 Objects  & $\Omega_0$ & $\Omega'_0$ & $\Omega''_0$ & $u_0$ & $v_0$ &  $A$ & $\sigma_0$  & $N_\mu +N_{vr}$ \\
Distance scale &  km s$^{-1}$  & km s$^{-1}$  & km s$^{-1}$ &
  km s$^{-1}$ & km s$^{-1}$ & km s$^{-1}$ &km s$^{-1}$ &  \\
    & kpc$^{-1}$ & kpc$^{-2}$ & kpc$^{-3}$ &
   &  & kpc$^{-1}$&  & \\
  \\[-7pt] \hline\\[-7pt]
OB~associations       & 30.03 & -4.56 & 1.04 & 6.53 & 11.46 & 17.10 & 6.8499 &  64+50\\
$r_{ph}=0.8\,r_{bh}$  & $\pm0.73$ & $\pm0.16$ & $\pm0.14$ & $\pm0.95$ & $\pm1.22$ &  $\pm0.60$ & &\\
 systematic errors   & -0.05 & 0.01 & -0.01 & -0.01 & -0.05 &  -0.03 & &\\
  \\[-7pt] \hline\\[-7pt]
OB~associations       & 29.57 & -4.20 & 0.72 & 8.52 & 8.57 & 15.75 & 7.5792  & 64+50 \\
$r_{tg}=1/\varpi_g$ & $\pm0.62$ & $\pm0.15$ & $\pm0.14$ & $\pm1.06$ & $\pm1.24$ &  $\pm0.56$ & & \\
systematic errors   & -0.02 & 0.01 & -0.01 & -0.01 & -0.06 &  -0.03 & &\\
  \\[-7pt] \hline\\[-7pt]
Field stars  & 28.70  & -4.20 & 0.88 & 6.72 &  9.78 & 15.75 & 13.0948  & 1899+913 \\
$r_{ph}=0.8\,r_{bh}$ & $\pm0.20$ & $\pm0.06$ & $\pm0.03$ & $\pm0.36$ & $\pm0.41$ &  $\pm0.23$ &  &\\
systematic errors   & 0.01 & 0.08 & -0.11 & 0.14 & -0.87 &  -0.30 & &\\
corrected values  & 28.69  & -4.28 & 0.99 & 6.58 & 10.65 & 16.05 &   &  \\
  \\[-7pt] \hline\\[-7pt]
Field stars  & 28.68  & -4.20 & 0.88 & 7.43 &  9.56 & 15.75 & 13.1919 & 1915+911\\
$r_{tg}=1/\varpi_g$ & $\pm0.18$ & $\pm0.05$ & $\pm0.02$ & $\pm0.36$ & $\pm0.40$ &  $\pm0.19$ &  &\\
systematic errors   & -0.20 & 0.17 & -0.13 & 0.33 & -0.97 &  -0.64 & &\\
corrected values  & 28.88  & -4.37 & 1.01 & 7.10 &  10.53 & 16.39 &   &\\
 \\[-7pt] \hline\\[-7pt]
\end{tabular}
\end{table*}
%---------------------------------------------------------------

We simulated the distribution of random errors in distance moduli,
$DM$,  and in trigonometric parallaxes, $\varpi_g$, to estimate the
biases  in the parameters of the rotation curve and the solar motion
towards the apex. The distribution of true distances is supposed to
be close to the observed distances. We calculated the true values of
the proper motions $\mu_l$ (Eq.~\ref{mu1}) and velocities $V_r$
(Eq.~\ref{vr1}) using  observational distances and the parameters
listed in Table 3 and added to them normally distributed errors with
the standard deviations $\varepsilon_{\mu t}$ and $\varepsilon_{vr}$,
respectively. We then  simulated "wrong" distances and solved the
systems of Eqs~\ref{mu1} and \ref{vr1} to determine the parameters of
the rotation curve and the solar motion. We repeated this procedure
10$^3$ times. The average shifts between the calculated and true
values of the parameters are also listed in Table 3. We can see that
the systematic corrections to the parameters obtained for the sample
of OB~associations  do not exceed $\sim10\%$ of the values of random
errors. Such small values of systematic errors are due to the great
accuracy of the relative distances (without consideration of the
distance-scale uncertainty) to OB~associations.

It follows from Table~3  that the  values of the parameters of the
Galactic rotation curve and the solar motion towards the apex,
$\Omega_0$, $\Omega'_0$, $\Omega''_0$, $u_0$ and $v_0$, derived with
photometric and trigonometric distances to OB~associations are
consistent within  the errors. The angular velocity of the Galactic
disk at the solar distance, $\Omega_0$, calculated for the two
distance scales has   the same values of $30.0\pm0.70$ km s$^{-1}$
kpc$^{-1}$. Such a good agreement  is due to the fact that both the
left-hand part of Eq.~\ref{mu1} ($4.74r\mu_l\cos b^{-1}$) and the
term with $\Omega_0$ in the right-hand part ($- \Omega_0 r \cos b$)
are proportional to the distance $r$, and hence distance-scale
changes have little effect on the inferred angular velocity
$\Omega_0$. Note that \citet{bobylev2019} obtained a similar value of
$\Omega_0$ equal to $\Omega_0=29.7\pm0.1$ km s$^{-1}$ kpc$^{-1}$ from
an analysis of the {\it Gaia} DR2 data for a sample of OB~stars.

We also derived the parameters of the rotation curve and the solar
motion from the kinematics of high luminosity field stars. The
parameters $\Omega_0$, $\Omega'_0$, $\Omega''_0$, $u_0$ and $v_0$
were also calculated for two sets of distances (Table~3). Here the
weight factors (Eqs~\ref{pvl} and~\ref{pvr}) were computed with the
use of the uncertainties $\varepsilon_{\mu l}$ and $\varepsilon_{vr}$
of measurements of proper motions and line-of-sight velocities of
individual stars. We excluded from consideration the velocities of
objects deviating more than 40 km s$^{-1}$ from the rotation curve,
so the numbers of conditional equations ($N_\mu+N_{vr}$) are a bit
different for photometric (1899+913) and trigonometric (1915+911)
sets of distances.

Table 3 also lists  the systematic errors and corrected parameters of
the rotation curve and the solar motion calculated for the sample of
field stars. Here the systematic errors appear to be comparable to
the random errors, so we corrected the calculated values of the
parameters for the systematic shift.

We can see that the parameters derived for two sets of distances
to field stars agree within the errors. Moreover, the parameters
obtained for OB~associations and field stars are consistent within
the errors.

Figure~\ref{rot} shows two rotation curves of the Galaxy  and the
azimuthal velocities of OB~associations calculated for photometric
and trigonometric distances to OB~associations. The corresponding
values of the rotation velocity at the solar distance are
$\Theta_0=225$ and 222 km s$^{-1}$, respectively. We can see that the
two rotation curves are practically flat in the 3-kpc solar
neighborhood. On the whole, the differences in the two rotation
curves can be thought to be insignificant.

We also calculated the parameters of the rotation  curve and the
solar motion towards the apex for the solar Galactocentric distance
$R_0=8.2$ kpc \citep{gravity2019}. For the set of photometric
distances to OB~associations, we obtained the following values:
$\Omega_0=30.09\pm0.73$ km s$^{-1}$ kpc$^{-1}$,
$\Omega'_0=-4.13\pm0.15$ km s$^{-1}$ kpc$^{-2}$,
$\Omega''_0=0.85\pm0.12$ km s$^{-1}$ kpc$^{-3}$, $u_0=6.75\pm0.94$ km
s$^{-1}$, $v_0=11.86\pm1.21$ km s$^{-1}$ and $A=16.95\pm0.62$ km
s$^{-1}$kpc$^{-1}$, which all except $\Omega'_0$ agree well with
those computed for $R_0=7.5$ kpc (first row of Table~3). We can see
that the value of $\Omega'_0$ decreases with increasing $R_0$, but
the value of the Oort constant $A=-0.5 \Omega'_0 R_0$ remains nearly
the same. The rotation curve obtained for $R_0=8.2$ kpc is also
nearly flat, but the value of $\Theta_0$ amounts to 247 km s$^{-1}$
here.

Here we supposed that the centroid of OB~associations rotates with
the velocity, $v_\varphi$, which is nearly equal to the velocity of
the rotation curve, $v_c$. The difference between them,
$v_\varphi-v_c$, the so-called asymmetric drift, is determined by the
Jeans equation and can be estimated from the following formula:

\begin{equation}
v_\varphi-v_c= \frac{\overline{\sigma_R^2}}{80} \;\; \textrm{km
s}^{-1},\label{ad}
\end{equation}

\noindent where $\sigma_R$ is the radial velocity dispersion of the
disk subsystem considered  \citep{binney2008}. For the sample of OB
associations, we adopted the value of $\sigma_R=9$ km s$^{-1}$ and
found  the asymmetric drift to be $v_\varphi-v_c \approx 1$ km
s$^{-1}$, which corresponds to the uncertainty of 0.1 km
s$^{-1}$kpc$^{-1}$ in the value of $\Omega_0$, what amounts to  only
18\% of its random error (Table 3). So the centroid of
OB~associations can be thought to rotate with the velocity of the
rotation curve.

Table~3 shows that the solar azimuthal velocity determined with
respect to the centroid of OB~associations, $v_0$, lies in the range
8--12 km s$^{-1}$ which is consistent with the values obtained in
other studies \citep{schonrich2010,tian2015,bobylev2018}.

\subsection{Residual velocities of OB~associations in the Galactic plane}

Residual velocities are the observed heliocentric velocities
corrected for the Galactic rotation and the solar motion towards the
apex: $V_{res}=V_{obs}-V_{rot}-V_{ap}$. The residual velocities show
how well objects  follow the Galactic rotation law and are indicators
of non-circular motions. In this section we consider residual
velocities  in the Galactic plane directed in the radial and
azimuthal directions. The radial component of the residual velocity,
$V_R$, is directed along the Galactic radius-vector  and its positive
value corresponds to the motion away from the Galactic center while
the azimuthal component, $V_T$, is tangent to circular orbits and its
positive value corresponds to an additional velocity in the sense of
Galactic rotation. Note that residual velocities are nearly
independent on the choice of the solar Galactocentric distance in the
range 7--9 kpc.

\begin{figure*}
\resizebox{\hsize}{!}{\includegraphics{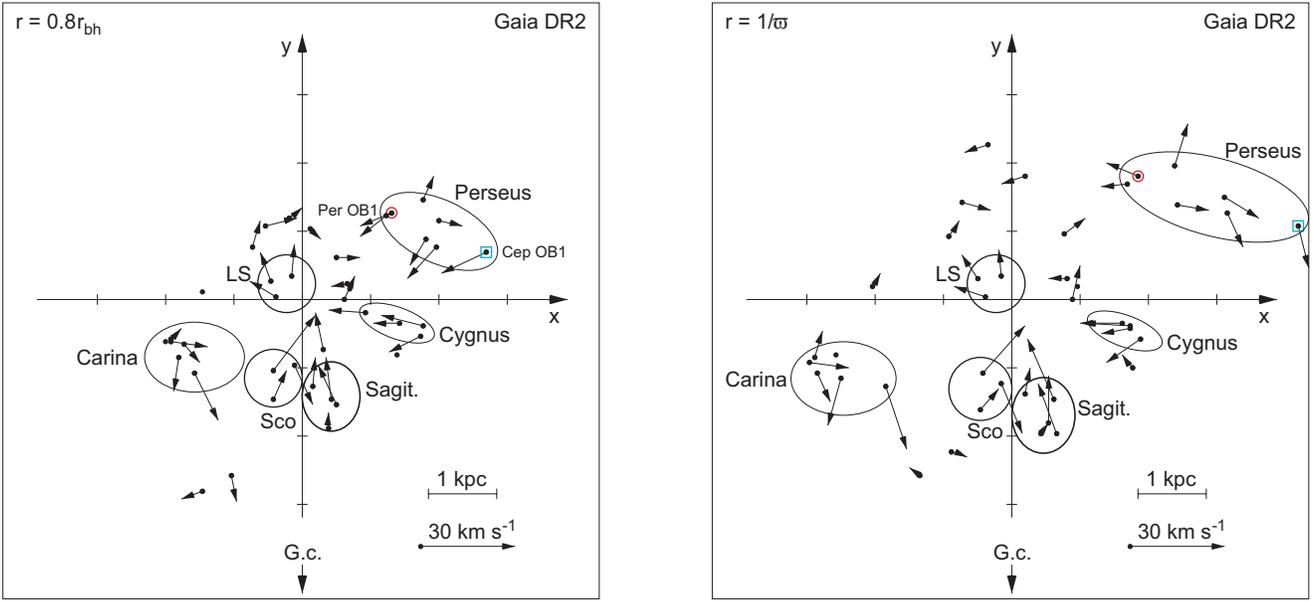}}
\caption{Distribution of the residual velocities of OB~associations
calculated with  photometric and trigonometric distances.  We present
only associations with $n_{vr}\ge 5$ and $n_\mu\ge 10$. Associations
with small residual velocities ($|V_R|<3$ and $|V_T|<3$ km s$^{-1}$)
are shown by  black circles without any vector. The Per OB1
association (marked by the red circle) and the Cep OB1 association
(marked by the blue square) have the residual velocities $V_R$ (Per
OB1) and $V_T$ (Cep OB1) depending most strongly on the choice of the
distance scale. Also show are the boundaries of the Sagittarius,
Scorpio, Carina, Cygnus, Local System and Perseus star-gas complexes.
The $x$-axis is directed in the sense of Galactic rotation and the
$y$-axis points away from the Galactic center. The Sun is at the
origin. The Galactic center (G.~c.) is in the bottom.}\label{resass}
\end{figure*}

Figure~\ref{resass} shows the distribution of the residual
velocities of OB~associations in the Galactic plane calculated
with photometric and trigonometric distances. The residual
velocities computed with the photometric distance scale are
determined with respect to the rotation curve calculated with
photometric distances and vice versa (Table~3).
Figure~\ref{resass} shows only OB~associations with the median
velocities derived from at least 10 proper motions ($n_\mu\ge 10$)
and 5 line-of-sight velocities ($n_{vr}\ge 5$) of member stars.
The root-mean-square differences, $\Delta V_R$ and $\Delta V_T$,
between residual velocities calculated with trigonometric and
photometric distances amount to $\Delta V_R=3.6$ and $\Delta
V_T=5.8$ km s$^{-1}$, respectively. The residual velocities of the
Per OB1 and Cep OB1 associations appear to be the most sensitive
to the choice of the distance scale. The radial residual velocity,
$V_R$, of the Per OB1 association changes by 8 km s$^{-1}$: from
$V_R=-6.6$ km s$^{-1}$ ($r_{ph}=1.83$ kpc) to 1.7 km s$^{-1}$
($r_{tg}=2.58$ kpc), which corresponds to the greatest change in
the velocity $V_R$ among OB~associations considered. The Cep OB1
association demonstrates the greatest change in the azimuthal
residual velocity $V_T$: from $V_T=-11.6$ km s$^{-1}$
($r_{ph}=2.78$ kpc) to +8.9 km s$^{-1}$ ($r_{tg}=4.32$ kpc), i.e.
by 21 km s$^{-1}$.

Figure~\ref{resass} also shows the boundaries of the Sagittarius,
Scorpio, Carina, Cygnus, Local System and Perseus star-gas
complexes identified by \citet{efremov1988}. A comparison of the
residual velocities calculated for the two  distance scales
suggests that the greatest changes take place in the Perseus
complex. In the photometric distance scale the majority of
OB~associations in the Perseus complex have the radial velocity
$V_R$ directed towards the Galactic center while in the
trigonometric distance scale their velocities $V_R$ are close to
zero. Note that the direction of the radial residual velocities in
the Perseus complex  is the foundation for all models of the
Galactic spiral structure and the Galactic resonance rings (see
section 4).

\begin{figure*}
\resizebox{\hsize}{!}{\includegraphics{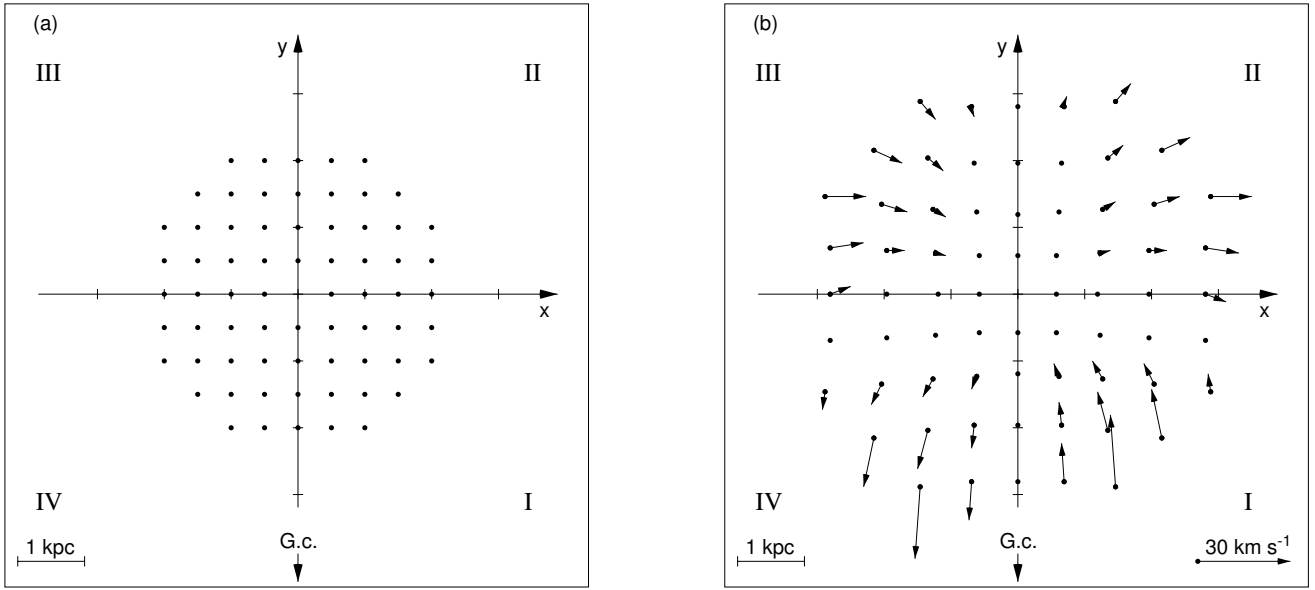}} \caption{Spurious
systematic motions emerging from the choice of the wrong distance
scale. (a) The initial distribution of the residual velocities of
test particles determined in the short distance scale of
OB~associations ($r_{ph}=0.8\,r_{bh}$). (b) The residual velocities
calculated with the use of the trigonometric distance scale,
$r_{tg}=(1.17/r_{bh}-0.11)^{-1}$. Roman numerals indicate the
quadrants. The radial component, $V_R$, of the spurious residual
velocities is directed towards the Galactic center in quadrants III
and IV and  away from it in quadrants I and II while the azimuthal
component, $V_T$, is directed in the sense of galactic rotation in
quadrants II and III and in the opposite sense in quadrants I and IV.
Particles with small residual velocities ($|V_R|<3$ and $|V_T|<3$ km
s$^{-1}$) are shown by  black circles without any vector.  The
$x$-axis is directed in the sense of Galactic rotation and the
$y$-axis points away from the Galactic center. The Sun is at the
origin. The Galactic center is in the bottom.} \label{resmod}
\end{figure*}

Figure~\ref{resmod} illustrates the appearance of systematic stream
motions due to the choice of a wrong distance scale. We scattered
test particles randomly over the galactic disk within 3.5 kpc from
the solar position and assigned to them  the velocities corresponding
to the Galactic rotation law which means that the residual velocities
equal zero (Fig.~\ref{resmod}a). For simplicity we adopted the flat
rotation curve with the angular velocity at the solar distance equal
to $\Omega_0=30$ km s$^{-1}$ kpc$^{-1}$. Let us suppose that we do
not know the true distances, corresponding to the short distance
scale ($r_{ph}=0.8\,r_{bh}$) but use instead of them the distances
corresponding to the long distance scale established by {\it Gaia}
DR2 parallaxes: $r_{tg}=(1.17/r_{bh}-0.11)^{-1}$ (Eq.~\ref{picor}).
Figure~\ref{resmod}(b) shows how the wrong distance scale affects the
residual velocities. The use of the trigonometric distance scale
causes the  appearance of spurious residual velocities which are
absent in the photometric distance scale.

Figure~\ref{resmod} shows that the spurious residual velocities
are very small in the vicinity of 1 kpc from the Sun ($|V_R|<3$
and $|V_T|<3$ km s$^{-1}$). However, objects located at distances
2--3 kpc from the Sun demonstrate significant (10--20 km s$^{-1}$)
spurious residual velocities.  In quadrant II, where the Perseus
complex is located, the spurious systematic motions are directed
away from the Galactic center and in the sense of Galactic
rotation. Generally, the radial component, $V_R$, of the spurious
residual velocities is directed towards the Galactic center in
quadrants III and IV and  away from it in quadrants I and II,
whereas the azimuthal component, $V_T$, is directed in the sense
of Galactic rotation in quadrants II and III and in the opposite
sense in quadrants I and IV. Note that the detection of a similar
picture in the distribution of residual velocities can suggest a
need to shrink the distance scale.

%----------------------- Figure 4  --------------------------
%-------------------------------------------------------------

%----------------------- Figure 5  ---------------------------
%----------------------------------------------------------------

\subsection{Motion in the Z-direction}

The residual velocities of OB~associations  in the direction
perpendicular to the Galactic plane, $V_z$, are determined with
the use of both  proper motions along Galactic latitude, $\mu_b$,
and  line-of-sight velocities, $V_r$:

\begin{equation}
V_z = 4.74 \mu_b \, r \cos b  + V_r \sin b + w_0, \label{vz}
\end{equation}

\noindent where $w_0$ is the velocity of the Sun in the
$Z$-direction.

Eq.~(\ref{vz}) indicates that the first term ($4.74 \mu_b r \cos
b$) depends on the distance $r$, and hence  the uncertainties in
distances can create spurious motions in $Z$-direction.  Two
conditions must be fulfilled for their appearance: objects must
not lie precisely in the Galactic plane ($b\ne 0$) and the
distance scale must be wrong.

If objects are distributed symmetrically with respect to the
Galactic plane then a wrong distance scale does not give rise to
systematic motions: objects lying above and below the Galactic
plane must acquire the additional velocities $V_z$  in opposite
directions which causes only an increase in the velocity
dispersion.

However, the  gas disk in the Galaxy  is rippled and young stars
born in it often lie 50--100 pc above or below the Galactic plane.
For example, all associations in the Cyngus complex (Cyg OB1, Cyg
OB3, Cyg OB8, Cyg OB9) are located above the Galactic plane
($b=+1..+3^\circ$) (see Table 2), which corresponds to the shift
of $\Delta z= 50$--130 pc above the plane. Here we adopted the
position of the Sun  with respect to the Galactic plane to be
$z_0=20$ pc. On the contrary, the  Per OB1  association is located
below the Galactic plane ($b=-3.14^\circ$) being shifted at
$\Delta z=-80$ pc with respect to the plane.

A ripple on the Galactic gas disk  and a wrong distance scale can
give rise to spurious systematic motions in the direction
perpendicular to the Galactic plane. The root-mean-square
difference between the velocities $V_z$ calculated with
photometric and trigonometric distances is 2.1 km s$^{-1}$. To
illustrate the emergence of spurious systematic motions in the
$Z$-direction we list the residual velocities $V_z$ for several
OB~associations determined for the two distance scales. Table~4
presents the Galactic coordinates, distances and residual
velocities $V_z$ obtained for the Per OB1, Cyg OB3, Cep OB1 and
Cep OB2 associations located above or below the Galactic plane. We
can see that the velocities $V_z$ derived for the photometric and
trigonometric distance scales differ, on average, by 2 km s$^{-1}$
but the danger is that this effect is systematic.

We solved the system of equations (\ref{vz}) with respect to the
solar velocity in the $Z$-direction, $w_0$,  for 50 OB~associations
with median velocities derived from at least 5 proper motions
($n_\mu\ge 5$) and 5 line-of-sight velocities ($n_{vr}\ge 5$) of
member stars to obtain  the values of $w_0=7.15\pm0.45$ and
$9.26\pm0.63$ km s$^{-1}$ for the photometric and trigonometric
distance scale, respectively. The corresponding velocity dispersions
in the vertical direction, $\sigma_z$, calculated for the two
distance scales have values of $\sigma_z=3.4$ and 4.4 km s$^{-1}$.

%The fact that the velocity dispersion $\sigma_z$ is greater in the
%trigonometric distance scale can also be due to the wrong distance
%scale.

Note that the analysis of the kinematics of high-luminosity field
stars yields very similar values of the solar vertical velocity:
$w_0=7.34\pm0.31$ km s$^{-1}$ (887 equations) and $8.43\pm0.32$ km
s$^{-1}$ (902 equations)  obtained for photometric and
trigonometric distance scales, respectively. We excluded field
stars with the residual velocities, $V_z$, greater than 40 km
s$^{-1}$. The vertical velocity dispersions of field stars derived
for the two distance scales are $\sigma_z=8.9$ and 9.2 km
s$^{-1}$, respectively.

%-------------------  Table 3  ---------------------------------
\begin{table*}
\centering \caption{Residual velocities $V_z$ of some
OB~associations}
 \begin{tabular}{lccccccc}
 \\[-7pt] \hline\\[-7pt]
 name & $l$  & $b$ & $r_{ph}$ & $r_{tg}$ & $\Delta z$ & $V_z$ ($r_{ph}$) & $V_z$ ($r_{tg}$)    \\
      &  deg & deg & kpc & kpc & kpc & km s$^{-1}$  & km s$^{-1}$   \\
  \\[-7pt] \hline\\[-7pt]
Cyg OB3   & 72.76  &   2.04 &   1.83 &   1.95 &  0.085 & 0.3 & 2.0 \\

Cep OB2   & 102.01 &   4.69 &   0.73 &   0.99 & 0.080 & 3.4 &  4.9 \\

Cep OB1   & 104.20 &  -0.94 &   2.78 &   4.32 & -0.026 & 0.0 &  -2.4 \\

Per OB1   & 134.70 & -3.14  & 1.83   &   2.58 & -0.080 & -0.9 & -3.1 \\

  \\[-7pt] \hline\\[-7pt]
\end{tabular}
\end{table*}
%---------------------------------------------------------------

\section{Discussion and conclusions}

We calculated the median parallaxes for 47 OB~associations
including at least 10 member stars with known {\it Gaia} DR2
parallaxes. The comparison of trigonometric and photometric
parallaxes to OB~associations revealed the  zero-point
displacement of {\it Gaia} DR2 parallaxes equal to $\Delta
\varpi=-0.11\pm0.04$ mas, which means that {\it Gaia} DR2
parallaxes are, on average, underestimated and distances derived
from them must be reduced.

\citet[][Table 1 there]{arenou2018} compared {\it Gaia} DR2
parallaxes with parallaxes of $\sim 6\times 10^4$ stars measured
by the Hipparcos satellite. The average visual magnitude  and the
zero-point offset of stars in their sample are
$\overline{G}=8.3^m$ and $\Delta \varpi=-0.118\pm0.003$ mas,
respectively. The average visual magnitude of stars of
OB~associations cross-matched with {\it Gaia} DR2 is
$\overline{G}=8.5^m$   and our value of the zero-point correction,
$\Delta \varpi=-0.11\pm0.04$, agrees  with the estimate by
\citet{arenou2018}.

Furthermore, the analysis of parallaxes of OB~associations and
high-luminosity stars in  field confirmed our previous conclusion
\citep{dambis2001,melnikdambis2009} that the distance scale to
OB~associations established by \citet{blahahumphreys1989} must be
reduced by 10--20\%.

We investigated how the choice of a wrong distance scale influences
the parameters of the rotation curve and found that the parameters
calculated with the use of  photometric and trigonometric distances
are consistent within the errors. In particular, the angular velocity
of the Galactic disk at the distance of the Sun, $\Omega_0$, computed
with the use of photometric and trigonometric distances to
OB~associations has the values of $30.0\pm0.7$ and $29.6\pm0.6$ km
s$^{-1}$ kpc$^{-1}$, respectively (Table~3).

The analysis of the residual velocities of OB~associations (i.e.
velocities corrected for the Galactic rotation and the solar motion
towards the apex) shows that they  depend strongly on the choice of
the distance scale. The root-mean-square  differences between the
residual velocities calculated with the use of photometric and
trigonometric distances  in projection on the Galactic radius vector,
azimuthal and vertical directions are $\Delta V_R=3.6$, $\Delta
V_T=5.8$ and $\Delta V_z=2.1$ km s$^{-1}$, respectively. A wrong
distance scale can   give rise to spurious systematic motions. The
distance scale determined by {\it Gaia} DR2 parallaxes creates
systematic motions  with the radial component, $V_R$, directed
towards the Galactic center in quadrants III and IV and away from it
in quadrants I and II and with  the azimuthal component, $V_T$,
directed in the sense of Galactic rotation in quadrants II and III
and in the opposite sense in quadrants I and IV. A discovery of a
similar velocity distribution can suggest the need to reduce  the
distance scale.

The residual velocities of objects located in the Perseus star-gas
complex appeared to be most sensitive to the choice of the distance
scale. In the case of the short photometric distance scale
($r_{ph}=0.8\,r_{bh}$) young stars of the Perseus complex demonstrate
conspicuous systematic motions in the direction toward the Galactic
center ($V_R=-6.7\pm2.7$ km s$^{-1}$), whereas in the case of the
trigonometric (uncorrected) distance scale these motions vanish
($V_R=-0.9\pm3.0$) being balanced by spurious systematic motions.

The  position of the density-wave spiral arms \citep{lin1964}
inside  the corotation circle (the radius at which the spiral
pattern rotates at the angular velocity equal to the angular
velocity of the Galactic disk) corresponds to the radial velocity
component directed toward the Galactic center ($V_R<0$).
\citet{lin1969} suggest that the Galactic spiral pattern rotates
with the angular velocity of $\Omega_s=13$ km s$^{-1}$ kpc$^{-1}$,
which puts the Perseus complex inside the corotation circle,
$\Omega_s<\Omega(R_\textrm{per}$). It is just the velocities
directed toward the Galactic center in the Perseus complex that
are the foundation for the concept of the Galactic spiral
structure and its modification for the four-armed spiral pattern
\citep{burton1974, georgelin1976, russeil2003, rastorguev2017,
bobylev2018,vallee2019}.

Another model of the Galaxy includes the bar and a two-component
outer ring $R_1R_2$ rotating with the angular velocity of the bar
$\Omega_b\approx50$ km s$^{-1}$ kpc$^{-1}$. Here also  the
direction of the residual velocities $V_R$ in the Perseus complex
is of great importance: their  direction toward the Galactic
center suggests the location of the Perseus region in the outer
resonance ring $R_2$
\citep{melnikrautiainen2009,melnik2011,rautiainen2010,
melnik2015,melnik2016,melnik2019}.

\section{Acknowledgements}

We  thank the anonymous referee  and the Editor for useful remarks
and suggestions. This work has made use of data from the European
Space Agency (ESA) mission {\it Gaia}
(https://www.cosmos.esa.int/gaia), processed by the  {\it Gaia} Data
Processing and Analysis Consortium (DPAC,
https://www.cosmos.esa.int/web/gaia/dpac/consortium). Funding for the
DPAC has been provided by national institutions, in particular the
institutions participating in the {\it Gaia} Multilateral Agreement.
This research has made use of the VizieR catalogue access tool, CDS,
Strasbourg, France. The original description of the VizieR service
was published  by \citet{ochsenbein2000}. A.~K. acknowledges the
support from the Russian Foundation for Basic Research (project nos.
18-02-00890 and 19-02-00611).

\end{document}